\documentclass{article}

\usepackage{PRIMEarxiv}

\usepackage[utf8]{inputenc} 
\usepackage[T1]{fontenc}    
\usepackage{hyperref}       
\usepackage{url}            
\usepackage{booktabs}       
\usepackage{amsfonts}       
\usepackage{nicefrac}       
\usepackage{microtype}      
\usepackage{lipsum}
\usepackage{fancyhdr}       
\usepackage{graphicx}       
\graphicspath{{media/}}     

\usepackage{amsmath}
\usepackage{multirow}
\usepackage{adjustbox}
\usepackage{tabularx}

\title{Improved methodology for deep aquifer characterization using hydrogeological, self-potential, and magnetotellurics data
}

\author{
  Young-Ho Seo\\
  Department of Civil and Environmental Engineering \\ Water Resources Research Center\\
  University of Hawaii at Manoa \\
  Honolulu, Hawaii 96822, USA \\
  \texttt{yhseo@hawii.edu} \\
   \And
  Aly I. El-Kadi \\
  Department of Earth Sciences \\ Water Resources Research Center \\
  University of Hawaii at Manoa \\
  Honolulu, Hawaii 96822, USA\\
  \And
  Niels Grobbe\\
  Hawai‘i Institute of Geophysics and Planetology,\\ School of Ocean and Earth Science and Technology and Water Resources Research Center\\  University of Hawaii at Manoa \\
  Honolulu, Hawaii 96822, USA \\
  \And
  Jonghyun Lee\\
  Department of Civil and Environmental Engineering \\ Water Resources Research Center\\
  University of Hawaii at Manoa \\
  Honolulu, Hawaii 96822, USA \\
  \texttt{jonghyun.harry.lee@hawaii.edu} \\
}

\begin{document}
\maketitle
\begin{abstract}
Estimation of subsurface properties such as hydraulic conductivity using hydrogeological data alone such as hydraulic head and well core data is typically challenging with sparsely available wells in field sites. Geophysical data, such as Self-potential (SP) and Magnetotelluric (MT) data, measured at the ground surface and/or in boreholes, can be acquired additionally to improve our understanding of the underlying hydrogeological structure and aid in the interpolation of data between wells, and possibly the extrapolation of data. However, in order to determine hydraulic conductivity, it is necessary to identify a proper petrophysical relationship between hydraulic conductivity and the inferred geophysical properties, such as electrical conductivity. Such relationships may not exist, may not yet be derived, or it may not be unique. 

In this work, we propose a joint-inversion approach without any assumptions about the petrophysical relationship. The proposed framework makes use of self-potential data, which connects groundwater flow velocity to measured electrical potential differences, and magnetotelluric data (or, e.g., electrical resistivity tomography data), to simultaneously estimate hydraulic conductivity and electrical conductivity of the subsurface. 

A spectral method is used to solve the self-potential forward problem, allowing for accurate computation of the flow velocity field derivatives required by the groundwater flow equation. To accelerate the joint data inversion, high-dimensional hydraulic conductivity and electric resistivity fields are estimated by using a dimension reduction technique through the Principal Component Geostatistical Approach. The approach assures that the estimated parameters and their uncertainties are quantified within only a few hundred coupled forward model runs. 

To demonstrate the applicability and robustness of the proposed joint-inversion method, several inversion tests are performed, using hydro geophysical data sets generated from subsurface models of hydraulic conductivity and electrical conductivity, created with and without petrophysical relationships. The findings demonstrate that the joint hydraulic head-SP-MT data inversion can be reasonably used to estimate hydraulic conductivity and electrical resistivity, even in the absence of knowledge of a one-to-one petrophysical relationship. On average, the proposed joint inversion yields 25\% improvement in the hydraulic conductivity estimates relative to a single data-type inversion, i.e., using only of hydraulic head and core data from wells; the single data-type inversion approach can only identify the subsurface structure near the observation wells.
Furthermore, additional joint head-MT inversion tests were performed using the ``perfect'' petrophysical relationship which we use to create the synthetic true field in order to compare the proposed SP-augmented approach to the ``best scenario'' joint inversion where no uncertainty exists in the petrophysical relationship identification.

Our proposed joint inversion approach of head-MT-SP data, where the SP-data compensates for the fact that no known petrophysical relationship is used, provided close agreement with the joint inversion of head and MT data, which uses a known petrophysical relationship. The successful inversion tests show the usefulness of using SP data, which connects hydrogeological properties and geophysical data, in the absence of a petrophysical relationship.
\end{abstract}


\section{Introduction}
Hydraulic conductivity is one of the key parameters to understanding the groundwater flow patterns for various engineering applications, such as water resources management, contaminant transportation and seawater intrusion and accompanying salinization ~\cite{lee2016scalable,kang2017improved}. However, estimating hydraulic conductivity in complex, heterogeneous aquifers using only hydrogeological data from wells is challenging, primarily due to the limited number of well locations compared to the extent of the subsurface domain of interest, or in other words the model domain size. The subsurface inverse problem of a huge number of unknown hydraulic conductivity values using a few noisy well observations becomes ill-posed, which means that the inverse solution is non-unique or sensitive to observation error. To account for the uncertainty in the estimate, the inverse problem has been treated within a stochastic framework, rather than using the deterministic method that yields only a single best estimate~\cite{mclaughlin1996reassessment,carrera2005inverse,kaipio2007statistical,slater2007near,oliver2008inverse}. 

To overcome the ill-posedness originating in the limited number of observations, the geostatistical inversion approach~\cite{kitanidis1983geostatistical,kitanidis1995quasi} is widely used for various subsurface applications \cite{cirpka2000sensitivity,michalak2004estimation,cardiff20113, cardiff2012field}. The approach uses spatial correlation of the underlying unknown field as prior information within the Hierarchical Bayesian framework \cite{kitanidis2012generalized,kitanidis2010bayesian}.  

Still, unless a large number of wells is available~\cite{hochstetler2016hydraulic}, hydrogeological data alone such as hydraulic head and borehole data at the wells only provides overly smooth, low resolution images of the subsurface with large uncertainty due to the diffusive nature of the groundwater flow equation. To address this issue, in the last decades, there have been various efforts to characterize the spatially variable hydraulic conductivity, using for example geoelectrical measurements~\cite{purvance2000geoelectric, slater2007near, chandra2008estimation, perdomo2014hydraulic}. Hydrogeophysical data, such as Electrical Resistivity Tomography (ERT), self-potential(SP), time-domain electromagnetics (EM), and magnetotellurics (MT) acquired either at the ground surface and/or in boreholes, can provide additional information on the subsurface hydrogeological structure, as well as serve as supplemental data for interpolation between the wells and possibly extrapolation beyond the wells. A number of studies have investigated the joint inversion using hydrogeological and geophysical data together to infer various subsurface parameters~\cite{jardani2009stochastic,huisman2010hydraulic, mboh2012coupled, jardani2013stochastic, kang2020improved}. However, in order to infer hydraulic conductivity, a suitable petrophysical relationship between hydraulic conductivity and the relevant geophysical model parameters, such as electrical conductivity, needs to be identified. Such a relationship may not be uniquely determined, or may even not exist.

To address the challenge of needing to identify suitable petrophysical relationships, we investigate the effectiveness of using self-potential and magnetotelluric data to inform the joint-inversion process for deep aquifer characterization. SP data is based on the measurement of electrical potential differences induced by groundwater flow~\cite{revil2013self,grobbe2021multi} and has been used to successfully estimate hydrogeological parameters \cite{jardani2009stochastic,straface2011joint, ozaki2014self}. Previous studies have investigated joint inversion for hydrogeological parameter-estimation using self-potential data, due to its advantage in relating the method's physics to the groundwater flow process~\cite{straface2011joint,jardani2013stochastic,soueid2014hydraulic, kang2020improved, han2022characterization}. For example, ~\cite{soueid2014hydraulic} applied the geostatistical approach to the inverse modeling of hydraulic conductivity and specific storage using hydraulic head and self-potential data. 

Magnetotellurics is a passive geophysical method for characterizing the electrical resistivity of the subsurface using the natural magnetic and electric field of Earth~\cite{chave2012magnetotelluric}. A number of geophysical studies of MT-inverse modeling for characterizing electrical conductivity have been conducted~\cite{unsworth2004electrical, avdeev20093d, aizawa2011temporal, meqbel2014deep, yang2015three, ledo20213d, yang2021electrical}. The typically low-frequencies being used and accompanying large wavelengths offer an advantage to the MT method: it allows for subsurface investigations of the resistivity in deeper regions up to the depth of a few kilometers. Incorporating MT data in the joint-inversion process can thus be expected to allow for characterization of the hydraulic conductivity and electrical resistivity for deep aquifers, for example, for geothermal heat energy exploitation~\cite{simmons2021interpretation}. To date, little research has been carried out on the study of estimating hydraulic conductivity explicitly using a joint inversion of MT, SP, and hydrogeological data sets. 

In this work, we propose a new joint inversion method for deep aquifer characterization that does not assume any petrophysical relationship, by incorporating MT, self-potential (SP), and hydrogeological data sets. In the proposed framework, hydraulic conductivity and electrical conductivity fields are simultaneously estimated through self-potential data fitting that links the groundwater velocity to the electrical conductivity. The self-potential forward problem is solved with a spectral method that allows for an accurate calculation of the derivatives of velocity fields as required in the governing equation. For computationally efficient site characterization, the Principal Component Geostatistical Approach (PCGA)~\cite{lee2014large,kitanidis2014principal} is utilized. The joint hydrogeophysics data inversion typically requires prohibitive computation and storage costs for Jacobian and dense covariance matrices especially for high-dimensional inverse problems with multi-physics coupled forward solvers. PCGA accelerates the geostatistical inversion by avoiding the computation of the Jacobian matrix product with the finite-difference approximation through the row-rank approximation of the prior covariance matrix. PCGA has already been applied to several large data set inverse problems, for example, estimating hydraulic conductivity using tracer concentration breakthrough data of synthetic cases~\cite{lee2014large}, reactive transport data~\cite{fakhreddine2016imaging}, temperature data~\cite{lee2018fast} and laboratory-scale sandbox experiments~\cite{lee2016scalable}. ~\cite{kang2020improved} conducted the joint inversion for hydraulic conductivity and DNAPL saturation by using the hydraulic head and partitioning trace data sets with self-potential as additional information. In this study, PCGA is implemented to estimate the high-dimensional hydraulic conductivity and electrical resistivity fields in synthetic heterogeneous aquifers and to quantify its estimation uncertainty, utilizing only a few hundred forward model runs.  

The remainder of the paper is organized as follows: in Section~\ref{sec:methods} we present an overview of the governing equations used in the joint data inversion. We then introduce the synthetic deep aquifer characterization configuration in Section~\ref{sec:applications}. We show how well the proposed method performs in comparison to the single data inversion using hydrogeological data or MT data alone. Lastly, the discussion and conclusions, including a reproducible code for the examples shown in the paper, are presented in Section~\ref{sec:conclusion}.

\section{Methods}
\label{sec:methods}
\subsection{Governing Equations}
For the forward problem, coupled governing equations for groundwater flow, SP, and MT are used to incorporate hydraulic head, self-potential, and magnetotellurics data sets into our proposed subsurface characterization framework.   
\subsubsection{Hydrogeology: Groundwater Flow}
The groundwater flow equation describing fluid flow through a porous medium at the Darcy scale is
\begin{equation}
    \label{eq:gw}
    \nabla \cdot \left[ \textbf{K}\left( \textbf{x} \right) \nabla h(\textbf{x},t) \right] = \textbf{S} \frac{\partial h(\textbf{x},t)}{\partial t}
\end{equation}
where $\textbf{x} = \left(x_1, x_2, x_3\right)$ is the spatial coordinate $\text{[m]}$, $\textbf{K} \left(\textbf{x}\right)$ is the hydraulic conductivity tensor $\text{[m/d]}$, $\textbf{S}$ is specific storage $\text{[1/m]}$, and $h$ is the hydraulic head $\text{[m]}$. Note that hydraulic conductivity estimation with the groundwater flow equation and hydraulic data alone typically results in overall smoothed outcome due to the widely used Gaussian (two-point correlation) assumption, as well as the associated estimation uncertainty becoming large far away from the observation wells ~\cite{lee2013bayesian,lee2014large}.

\subsubsection{Magnetotellurics}
Magnetotellurics (MT) is a passive geophysical method in which natural electromagnetic fields are used to image the electrical conductivity or resistivity of the subsurface. ~\cite{key2016mare2dem} developed the MARE2DEM code for 2-D anisotropic forward and inverse modeling of Magnetotellurics (MT) data and frequency-domain controlled-source electromagnetic (CSEM) data acquired from geophysical surveys. We use  MARE2DEM for the forward modeling and consider a 2D electrical conductivity model to generate synthetic MT observations. 
The governing equations for the frequency-domain electromagnetic field are shown below, assuming an $e^{-i\omega t}$ time-dependence, with angular frequency $\omega$:
\begin{equation}
    \label{eq:MT1}
    \nabla \times \mathbf{E}  - i \omega \mu \mathbf{H} = \mathbf{M}_{s} 
\end{equation}

\begin{equation}
    \label{eq:MT2}
    \nabla \times \mathbf{H} - \sigma \mathbf{E} = \mathbf{J}_{s}
\end{equation}
where $\mathbf{E}$ is the frequency-domain electric field [V/m], $\mathbf{H}$ is the magnetic field [A/m], $\mathbf{M}_{s}$ is a magnetic current source [V/m$^2$], and $\mathbf{J}_{s}$ is an electric current source [A/m$^2$]. $\mu$ denotes the magnetic permeability and $\sigma$ the complex electrical conductivity (for most low-frequency geophysical applications, one can neglect the imaginary component containing the dielectric permittivity term, in the so-called quasi-static approximation). After the Fourier transformation, the governing equations can be expressed in a compact form:

\begin{equation}
    \label{eq:compact}
    -\nabla \cdot \left( A \nabla \mathbf{u} \right) + C \mathbf{u} = \mathbf{f} \quad \rm in \:\: \Omega \quad \quad \mathbf{u} = \mathbf{0} \quad \rm on \:\: \partial \Omega
\end{equation}

\begin{equation}
    \label{eq:MTsource}
    \mathbf{f} = \nabla \cdot (AQ^{T}\mathbf{s}_{t}) - \mathbf{s}_{x}
\end{equation}
where
\begin{equation}
    R = \begin{pmatrix}
    0 & -1\\
    1 & 0
    \end{pmatrix}, \quad
    Q = \begin{pmatrix}
    0 & R\\
    R & 0
    \end{pmatrix}, \quad
    \mathbf{s}_{t} = (\hat{\mathbf{J}}^{s}_{t}, \hat{\mathbf{M}}^{s}_{t}),\quad 
    \mathbf{s}_{x} = (\hat{J}^{s}_{x}, \hat{M}^{s}_{x})
\end{equation}
where $\Omega$ is the model domain and $\mathbf{u} = \left(\hat{E}_{x}, \hat{H}_{x} \right)$. $x$ and hat $\left(\ \hat \ \ \right)$ denote the strike direction and the quantity in the wavenumber domain respectively. $\mathbf{f}$ is the source term and $t$ denotes transverse direction. The coefficient matrices $A$ and $C$ for MT are in \cite{key2016mare2dem}:

\begin{equation}
    \label{MT:coeff}
    A = \begin{pmatrix}
    \lambda \sigma_{t} & 0 \\
    0 & i \omega \mu \lambda' 
    \end{pmatrix}, \quad
    C = \begin{pmatrix}
    \sigma_{x} & 0 \\
    0 & i \omega \mu 
    \end{pmatrix},
    \nonumber
\end{equation}
where
\begin{equation}
    \sigma_{t} = \begin{pmatrix}
    \sigma_{y}  & 0 \\
    0 & \sigma_{z} 
    \end{pmatrix}, \quad
    \lambda^{-1} = \begin{pmatrix}
    -i\omega\mu\sigma_{y}  & 0 \\
    0 & -i\omega\mu\sigma_{z} 
    \end{pmatrix}, \quad
    \lambda'=R^{T}\lambda R 
\end{equation}
The numerical solutions of Equation \ref{eq:compact} provide the electric and magnetic fields of the strike direction. The transverse electric and magnetic fields can be obtained by:

\begin{equation}
    \mathbf{u}_{t}= (\hat{\mathbf{E}}_{t},\hat{\mathbf{H}}_{t}) = QA\nabla \mathbf{u}+Q^{T}AQ \mathbf{s}_t
\end{equation}

\subsubsection{Self-Potential}
The self-potential method is a passive geophysical method that measures the electrical potential differences to investigate dynamic subsurface processes, such as groundwater flow or geochemical reactions~\cite{revil2013self,GrobbeBardeCabusson2019,BardeCabussonetal2021,Reviletal2023}. Here, we focus on groundwater flow related self potential signals. The governing equation of self-potential can be derived, starting from the continuity equation for electrical charge~\cite{sill1983self}:
\begin{equation}
\label{eq:j}
\nabla\cdot\mathbf{j}=0
\end{equation}
where $\mathbf{j}$ is the current density (A/$\rm m^{2}$). The current density due to groundwater flow in a heterogeneous porous medium can represent the total flux of natural electrical charges~\cite{revil2007electrokinetic,boleve2007new,jardani2007tomography}:

\begin{equation}
\label{eq:j2}
\mathbf{j}=-\sigma\nabla\varphi+\hat{Q}_{v}\mathbf{q}
\end{equation}
where $\sigma$ denotes the electrical conductivity (S/m), $\varphi$ is the electrical potential (V), $\hat{Q}_{v}$ is the effective charge density ($\text{C}/\text{m}^{3}$), and $\mathbf{q}$ represents the specific discharge vector (m/s).
Then the self-potential ($\varphi$) equation can be obtained from Equations~\ref{eq:j} and~\ref{eq:j2}~\cite{kang2020improved}:
\begin{equation}
\nabla \cdot (\sigma\nabla\varphi)=\nabla \cdot (\hat{Q}_{v} \mathbf{q})
\label{eq:sp}
\end{equation}
with the boundary conditions: 
\begin{equation}
\begin{split}
\varphi=0 \quad \text{at} \quad \Gamma_{\text{D}} \\ -\mathbf{n} \cdot (\sigma \nabla \varphi - \hat{Q}_{v} \mathbf{q}) = 0 \quad \text{at} \quad \Gamma_{\text{N}} 
\end{split}
\end{equation}
The self-potential equation in Equation~\ref{eq:sp} relates the electrical conductivity $\boldsymbol \sigma$ (i.e., $\frac{1}{\rho}$) from Maxwell's equations (i.e., Equations~\ref{eq:MT1} and~\ref{eq:MT2}) to the groundwater discharge $\mathbf{q}=-\textbf{K}\left( \textbf{x} \right) \nabla h(\textbf{x},t)$ from the groundwater flow equation in Equation~\ref{eq:gw} without any petrophysical relationship. The unknown effective charge density $\hat{Q}_{v}$ can be inferred from the hydraulic conductivity using the empirical relationship identified in previous studies~\cite{jardani2007tomography,revil2007electrokinetic,jougnot2012derivation}, which will be explained in the next section. 

\subsection{Fast and Scalable Inverse Modeling: Principal Geostatistical Approach}
When solving the inverse problem, for example, using the geostatistical approach, the coupled governing equations need to be solved multiple times. For example, several tens of thousands of numerical simulations are required to compute the Jacobian and its products with the dense prior covariance matrix ~\cite{ghorbanidehno2020recent}.
To address the computational challenges associated with such inverse problems, the PCGA (Principal Component Geostatistical Approach) can be implemented as a ``matrix-free'' geostatistical inverse modeling approach that can avoid the direct calculation of the Jacobian matrix and its products by utilizing the principal components (low-rank approximation) of the prior covariance through a finite-difference approximation~\cite{lee2014large,lee2016scalable}. Here we briefly explain the main idea of PCGA. The quasi-linear geostatistical approach~\cite{kitanidis1995quasi} starts from the observation equation:
\begin{equation}
    \label{linear_eq}
    \mathbf{y} = \mathbf{h}(\mathbf{s})+\mathbf{v}
\end{equation}
where $\mathbf{y}$ is the $n \times 1$ vector of the observations, $\mathbf{s}$ is the $m \times 1$ vector of the unknown spatially distributed parameters such as hydraulic conductivity or resistivity, and $h$ is the forward model. The random variable $\mathbf{v}$ represents the error in the observations $y$ as well as the forward model $h$ and follows a Gaussian distribution with zero mean and covariance $\mathbf{R}$. The probability density function (pdf) of $\mathbf{y}$ given $\mathbf{s}$, the likelihood function, is following a Gaussian with mean $h(\mathbf{s})$ and covariance matrix $\mathbf{R}$. The prior of the unknown spatially distributed field $\mathbf{s}$ is parameterized as:
\begin{equation}
    \label{linear_eq2}
    \mathbf{s} =\mathbf{X} \boldsymbol{\beta}+\boldsymbol \varepsilon
\end{equation}
where $\mathbf{X}$ is a known $m \times p$ matrix representing deterministic trends/drifts, $\boldsymbol{\beta}$ is a vector consisting of $p$ unknown drift coefficients, and $\boldsymbol \varepsilon$ is the spatially correlated random variable with zero mean and the prior covariance matrix $\mathbf{Q}$ that is not explained with the drift/trend. Therefore, the prior $\mathbf{s}$ is a Gaussian with mean $\mathbf{X} \boldsymbol{\beta}$ and covariance matrix $\mathbf{Q}$. The posterior pdf $p''(s)$ can be obtained through Bayes' theorem, and the negative loglikelihood for the posterior pdf of $\mathbf{s}$ and $\boldsymbol{\beta}$ is minimized to obtain the maximum a posterior (MAP) or most likely value $\mathbf{\hat{s}}$:

\begin{equation}
    \label{eq:map}
    -\ln{p''(s,\boldsymbol{\beta})} = \frac{1}{2}\left(\mathbf{y}-h(\mathbf{s})  \right)^{\text{T}} \mathbf{R}^{-1} \left(\mathbf{y}-h(\mathbf{s}) \right) + \left(\mathbf{s}  - \mathbf{X} \boldsymbol{\beta}\right)^{\text{T}} \mathbf{Q}^{-1} \left(\mathbf{s}  - \mathbf{X} \boldsymbol{\beta} \right)
\end{equation}
One can update the latest estimation $\mathbf{\bar{s}}$ to a new solution until it converges to the most likely value $\mathbf{\hat{s}}$. The $n \times m$ Jacobian matrix $\mathbf{H}$, the derivative of $h$ with respect to $\mathbf{s}$ at $\mathbf{\bar{s}}$, can be expressed as:

\begin{equation}
    \label{Jacobian}
    \mathbf{H}=\left.\frac{\partial h}{\partial \mathbf{s}}\right|_{\mathbf{s}=\mathbf{\bar{s}}}
\end{equation}
Using the Jacobian matrix for the linearization of the forward model $h$,  the updated next solution can be evaluated as:

\begin{equation}
    \label{up_sol}
    \mathbf{\bar{s}}=\mathbf{X}\bar{\boldsymbol{\beta}}+\mathbf{QH}^{\text{T}}\bar{\xi}
\end{equation}
where the $n \times 1$ vector $\boldsymbol{\xi}$ and $p \times 1$ vector $\boldsymbol{\beta}$ can be computed by the following linear system:

\begin{equation}
\label{xiform}
\begin{bmatrix}
\mathbf{HQH^{\text{T}} + R} & \mathbf{HX}\\
(\mathbf{HX})^{\text{T}} & \mathbf{0}
\end{bmatrix}
\begin{bmatrix}
\boldsymbol{\bar{\xi}}\\
\boldsymbol{\bar{\beta}}
\end{bmatrix}
= 
\begin{bmatrix}
\mathbf{y}-h(\mathbf{\bar{s}})+\mathbf{H\bar{s}}\\
\mathbf{0}
\end{bmatrix}
\end{equation}
By repeating Equations~\ref{Jacobian}-~\ref{xiform} until the convergence of $\bar{\mathbf{s}}$, the best estimation $\hat{\mathbf{s}}$ can be obtained. After computing the best estimation, the posterior covariance can be used for the estimation of uncertainty. The posterior covariance of $\mathbf{s}$, $\mathbf{V}$, is the inverse of the Hessian of the objective function (Equation \ref{eq:map}) and can be simplified by using matrix identities:
\begin{equation}
\label{Cov1}
\mathbf{V} = \left( \mathbf{Q}^{-1}+\mathbf{H}^\text{T}\mathbf{R}^{-1}\mathbf{H}  \right)^{-1}
= - \mathbf{XM}+\mathbf{Q}-\mathbf{Q}\mathbf{H}^{\text{T}}\mathbf{\Lambda},
\end{equation}
where $\mathbf{X}$ and $\mathbf{\Lambda}$ can be obtained from:

\begin{equation}    
\label{cov2}
\begin{bmatrix}
\mathbf{HQH^{\text{T}} + R} & \mathbf{HX}\\
(\mathbf{HX})^{\text{T}} & \mathbf{0}
\end{bmatrix}
\begin{bmatrix}
\boldsymbol{\Lambda}^{\text{T}}\\
\mathbf{M}
\end{bmatrix}
= 
\begin{bmatrix}
\mathbf{HQ}\\
\mathbf{X}^{\text{T}}
\end{bmatrix}.
\end{equation}
For large-scale inverse problems such as deep aquifer characterization with multiple hydrogeological and geophysical data sets, the geostatistical approach would require a high computational cost for the construction and storage of the Jacobian matrix $\mathbf{H}$ and its products with the dense prior covariance matrix $\mathbf{Q}$, i.e., $\mathbf{HQ}$ and $\mathbf{HQH}^{\rm T}$. \cite{lee2014large} developed a scalable method called the principal component geostatistical approach (PCGA) that utilizes the low-rank of $\mathbf{Q}$ and a finite difference approach to avoid the direct construction of $\mathbf{H}$ and accurate approximation of the Jacobian-Covariance products. The low-rank approximation of $\mathbf{Q}$ can be assumed as:
\begin{equation}
\label{Q:approx}
\mathbf{Q} \approx \mathbf{Q}_K = \mathbf{ZZ}^T=\sum_{i=1}^{K}\zeta_{i}\zeta_{i}^T
\end{equation}
where $\mathbf{Q}_{K}$ is a rank-K approximation of $\mathbf{Q}$, $\mathbf{Z}$ is a $m \times K$ principal component matrix and $\zeta_{i}^T$ is \emph{i}th column vector of $\mathbf{Z}$, which is an $i$-th eigenvector of $\mathbf{Q}$ scaled by square root of its eigenvalue. The Jacobian-vector product $\mathbf{Hx}$ can be calculated by using the Taylor expansion of the forward model \emph{h}:

\begin{equation}
\label{Talyor}
h(\bar{\mathbf{s}}+\delta\mathbf{x}) = h(\bar{\mathbf{s}}) + \delta\mathbf{H}\mathbf{x}+\mathcal{O}(\delta^2)
\end{equation}
where $x$ is a $m \times 1$ vector and $\delta$ is the finite difference interval. When $\mathbf{x} = \mathbf{\bar{s}}$, a Jacobian-vector product $\mathbf{H\bar{s}}$ can be:

\begin{equation}
\label{eq:fd}
\mathbf{H\bar{s}} = \frac{1}{\delta}\left( h(\mathbf{\bar{s}}+\delta\mathbf{\bar{s}}) - h(\mathbf{\bar{s}}) \right) + \mathcal{O}(\delta) \approx \frac{1}{\delta}\left( h(\mathbf{\bar{s}}+\delta\mathbf{\bar{s}}) - h(\mathbf{\bar{s}}) \right)
\end{equation}
Similarly, the approximation of $\mathbf{HQ}$ can be computed by

\begin{equation}
\mathbf{HQ} \approx \mathbf{HQ}_{K} = \mathbf{H} \sum_{i=1}^{K}\zeta_{i}\zeta_{i}^{\rm T} = \sum_{i=1}^{K}(\mathbf{H}\zeta_{i})\zeta_{i}^{\rm T} \approx \sum_{i=1}^{K} \boldsymbol{\eta}_{i}\zeta_{i}^{\rm T}
\end{equation}
where
\begin{equation}
\boldsymbol{\eta}_{i} = \mathbf{H}\zeta_{i} \approx \frac{1}{\delta} \left( h(\mathbf{s}+\delta\zeta_{i})-h(\mathbf{s}) \right)
\end{equation}

\subsection{Joint Hydrogeological and Geophysical Inversion Strategy without Petrophysical Relationship}

To identify subsurface properties beyond the well locations, geophysical surveys such as Magnetotelluric and seismic investigations are widely used. However, geophysical survey results typically require a petrophysical relationship to convert the estimated geophysical properties, such as electrical conductivity (or resistivity) or seismic velocity to hydraulic conductivity. To address this challenge, we propose a novel joint inverse modeling framework to estimate hydraulic conductivity and electrical resistivity simultaneously through the use of additional self-potential data sets. 

In our geostatistical inversion approach, the unknown subsurface properties are represented in an augmented vector $\mathbf{s}$ consisting of the log-hydraulic conductivity field $\ln \mathbf{K}$ and the log-resistivity field $\ln \boldsymbol \rho$ ($\rho$=$1/\sigma$) as in Equation~\ref{eq:joint_inversion}. Note that for the same mesh for the groundwater flow and MT simulations, $\mathbf{s}$ is a $2m$ by $1$ vector concatenating the $m$ by $1$ $\ln K$ and $\ln \boldsymbol \rho$ vectors. In the same manner, the covariance matrix $\mathbf{Q}$ embeds the covariance of the hydraulic conductivity field $\mathbf{Q}_{\ln{K}}$ and the covariance of the resistivity field $\mathbf{Q}_{\ln \rho}$ and we assume there is no correlation between the hydraulic conductivity and resistivity in the prior for simplicity. One might use a parameterized correlation model between $\mathbf{K}$ and $\boldsymbol \rho$ for better results. 

\begin{equation}
\label{eq:joint_inversion}
\mathbf{s} =
\begin{bmatrix}
\ln \mathbf{K}\\
\ln \boldsymbol{\rho} 
\end{bmatrix}, \quad
\mathbf{Q} = 
\begin{bmatrix}
\mathbf{Q}_{\ln k} & \mathbf{0}\\
\mathbf{0} & \mathbf{Q}_{\ln \rho}
\end{bmatrix}
\end{equation}

In the joint data inversion framework, we need to utilize three forward models that simulate the groundwater flow, the self-potential, and the MT responses. The groundwater model $\mathbf{h}_{gw}$ uses hydraulic conductivity $\mathbf{K}$ to simulate the hydraulic heads and compute its corresponding groundwater velocities. The MT model $\mathbf{h}_{MT}$ uses electrical conductivity $\boldsymbol \sigma$ or resistivity $\boldsymbol \rho$ distributions to produce the electromagnetic signals. The self-potential model is adopted here to link these two models and eliminate the need for an often unknown or uncertain petrophysical relationship, by simply satisfying the continuity equation of the self-potential in Equation~\ref{eq:sp} with the currently estimated groundwater velocity and resistivity fields through self-potential observations $\boldsymbol \varphi$ within the Bayesian framework as in Equation~\ref{eq:map}. 

The Jacobian matrix $\mathbf{H}$ contains the information from the three models; the first row of $\mathbf{H}$ represents the Jacobian matrix between the hydraulic conductivity field and the hydraulic head observations, the second row consists of the Jacobian matrix (i.e., sensitivity) between the hydraulic conductivity field and the SP observations $\mathbf{H}_{SP-GW}$ and the Jacobian matrix between the resistivity field and the SP observation $\mathbf{H}_{SP-MT}$, and the third row represents the Jacobian matrix of the resistivity and the MT forward models, respectively as 
\begin{equation}
\label{eq:joint_jac}
\mathbf{H} = 
\begin{bmatrix}
\mathbf{H}_{GW} & 0\\
\mathbf{H}_{SP-GW} & \mathbf{H}_{SP-MT}\\
0 & \mathbf{H}_{MT} 
\end{bmatrix}
\end{equation}
\begin{equation}
\mathbf{H}_{GW} = \frac{\partial \mathbf{h}_{GW}}{\partial \ln \mathbf{K}},\quad \mathbf{H}_{SP-GW} = \frac{\partial \mathbf{h}_{SP}}{\partial \ln \mathbf{K}}, \quad \mathbf{H}_{SP-MT} = \frac{\partial \mathbf{h}_{SP}}{\partial \ln \rho},\quad \mathbf{H}_{MT} = \frac{\partial \mathbf{h}_{MT}}{\partial \ln \rho}    
\end{equation}
The Jacobian-vector product can be computed separately within each numerical model through finite difference as in Equation~\ref{eq:fd} and can be subsequently merged for the construction of the entire Jacobian matrix $\mathbf{H}$.

\section{Description of the Numerical Experiments}
\label{sec:applications}
\subsection{Synthetic Site Domain}
In this section, we assess the performance of our proposed joint inversion for estimating the subsurface properties in a synthetic deep aquifer. Here, two cases are considered; the first case represents a configuration in which the true hydraulic conductivity and resistivity fields are correlated through a petrophysical relationship. In the other case, we generated the true hydraulic conductivity and resistivity fields independently from their own Gaussian random field parameters. The performance of the joint inversion model using groundwater, self-potential, and MT models and data sets is evaluated by comparing the results against those obtained from the single hydrogeological inversion, which are obtained by using the observed well-core and hydraulic-head data.

\begin{figure}[htp!]
    \centering
    \includegraphics[width=\textwidth]{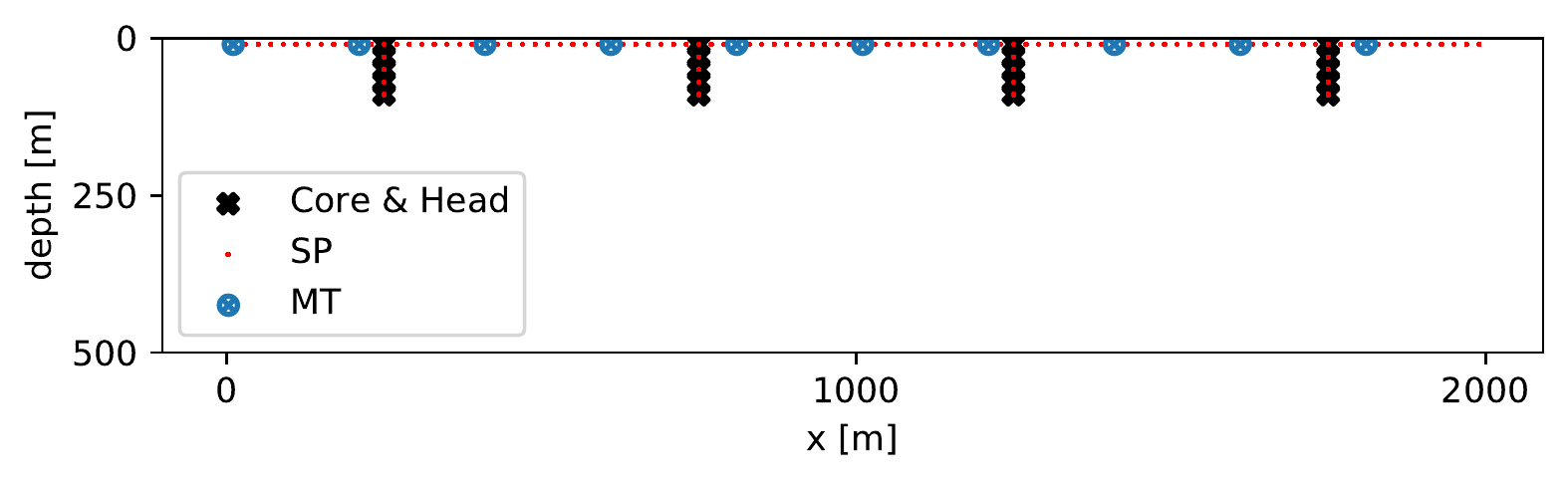}
    \caption{Measurement configuration}
    \label{fig:figure_scheme}
\end{figure}

As explained in the previous sections, we additionally utilize the self-potential model and sparse observations to simultaneously estimate hydraulic conductivity and resistivity fields without any assumed relationship between the two unknown fields. The domain and observation configurations for the inverse modeling tests are illustrated in Figure~\ref{fig:figure_scheme}.

The model domain is a 2D unconfined aquifer of 2 km by 2 km in which 4 observation wells provide core data for the hydraulic conductivity and hydraulic heads. The depth of the wells is 100 m, and we assume that the head and core data are measured at intervals of 20 m each. The self-potential data is measured at intervals of 20 m from the surface and wells. For the MT survey, 21 MT receivers over the 4 km surface (i.e., 4 km by 4 km domain extended from the 2 km by 2 km model domain) measure EM amplitude and phase for 32 frequency bands to estimate resistivity at depth in the inverse modeling. The relatively large-scale 2D modeling domains are intended to 1) evaluate how deep the proposed method can identify the properties and 2) eliminate the effect of boundary conditions so that the inversion can minimize the effect of wrongly assigned boundary conditions, which typically take place in practice.
\subsection{Subsurface Property and Data Generation} \label{Subsurface Property and Data Generation}
In order to evaluate the performance of our proposed method, we consider several hydraulic conductivity and resistivity fields. Specifically, we design the numerical experiments in terms of observation with two configurations: (1) the hydraulic conductivity and resistivity fields are determined through a known petrophysical relationship, and (2) the hydraulic conductivity and resistivity are determined independently. The purpose of these configurations is to check whether the proposed inversion method can perform well for deep aquifer characterization where one cannot uniquely determine the petrophysical relationship.
\begin{figure}[htp!]
    \centering
    \includegraphics[width=\textwidth]{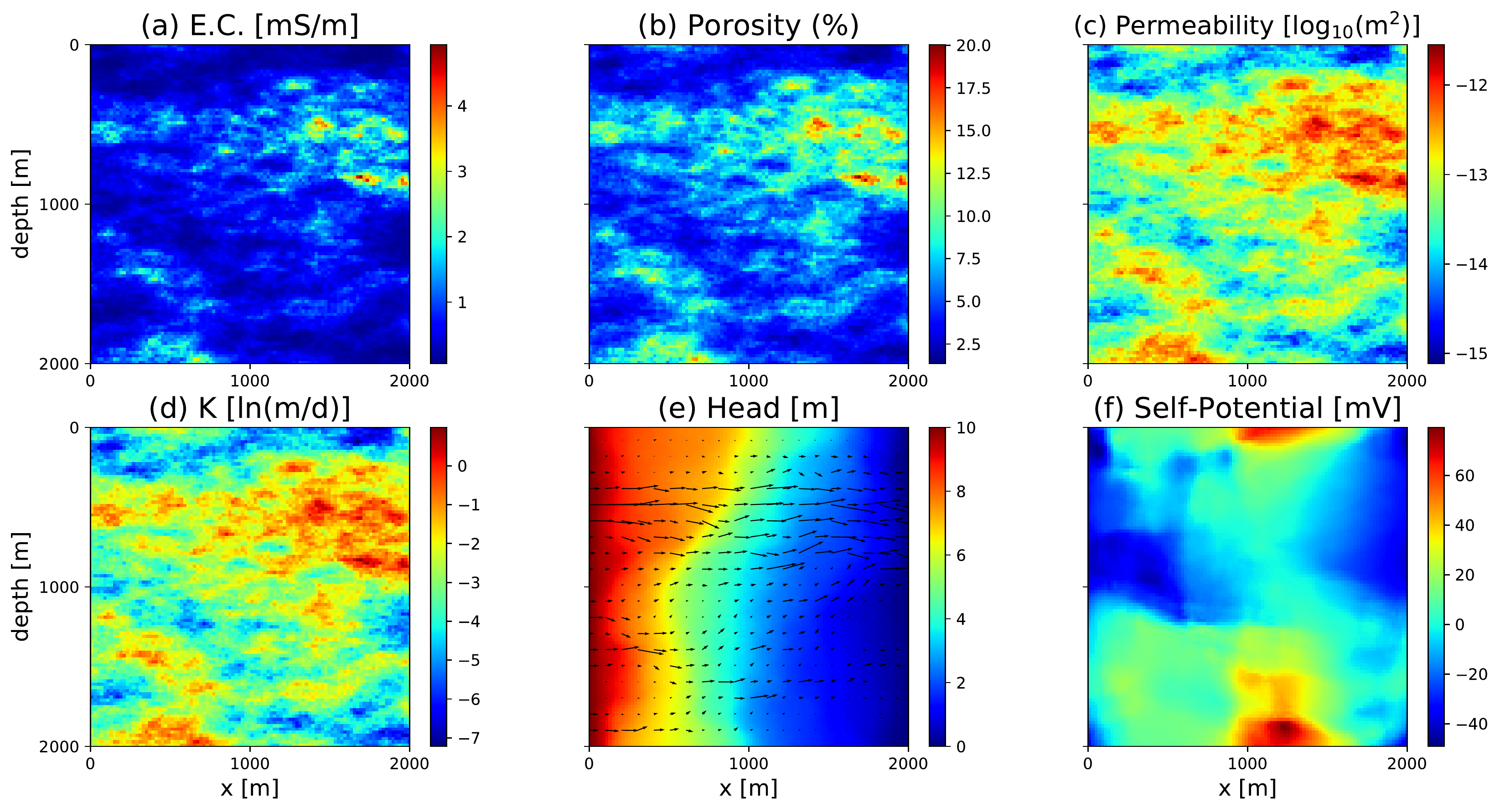}
    \caption{Data generation with petrophysical relationship; these fields are used for the inversion test as Case 1-1}
    \label{fig:figure_petro}
\end{figure}

The first configuration case assumes the known petrophysical relationship in the true field generation. For this, we first randomly generated the electrical conductivity $\sigma$, the reciprocal of the resistivity $\rho = \frac{1}{\sigma}$, from a log-normal distribution using Gaussian and exponential kernels with scale parameters as shown in Figure~\ref{fig:figure_petro} (a). Using the random field of the electrical conductivity, porosity is computed using Archie’s law as shown in Figure~\ref{fig:figure_petro} (b). Archie’s Law is an empirical expression for the electrical conductivity of porous media related to its porosity and fluid saturation of pores. In a saturated medium, Archie's Law can be reduced to~\cite{ewing2006dependence}: 
\begin{equation}
    \sigma = a \sigma_{b} \phi^{m}
\end{equation}
where $\sigma$ is the electrical conductivity of porous media, $\sigma_{b}$ is the electrical conductivity of the bulk brine solution, $\phi$ is the porosity and \emph{a} and \emph{m} are fitting parameters, respectively. The range of the parameters are chosen from the previous studies~\cite{cai2017electrical,thorslund2020global}. To convert porosity $\phi$ to the hydraulic conductivity, a modified Kozeny-Carman model~\cite{kozeny1927uber,carman1937fluid,costa2006permeability} is used with the Archie's equation that leads to:
\begin{equation}
\label{eq:phitok}
K = \frac{k \rho g}{\mu} =  C_{c} \frac{\phi^{m}}{(1-\phi)} \rho g
\end{equation}
where $k$ is the permeability,  $\rho$ is the density of the fluid, \emph{g} is the gravitational acceleration, $\mu$ is the viscosity, $C_{c}$ is a model factor, and $m$ is the Archie exponent. The permeabilities of different porous media were selected and modified from~\cite{costa2006permeability}. The resulting permeability field is shown in Figure~\ref{fig:figure_petro} (c) with a range of $10^{-16}$ to $10^{-12} \ \textrm{m}^{2}$. After generating the permeability field, the hydraulic conductivity is calculated as in Figure \ref{fig:figure_petro} (d) by its definition:
where. Using the hydraulic conductivity, the hydraulic head and its corresponding groundwater velocity were simulated as shown in Figure \ref{fig:figure_petro}(e) and observations were extracted at the data locations in Figure~\ref{fig:figure_scheme}. The self-potential model in Equation~\ref{eq:sp} requires the effective charge density. In this study, we following the previously developed empirical relationship ~\cite{jardani2007tomography,jougnot2012derivation,revil2017transport}:
\begin{equation}
\log_{10}{\hat{Q}_{v}} = -9.2 -0.82 \log_{10}{k_{eff}}
\end{equation} 

\section{Results}
\label{sec:results}
\subsection{Case 1: Inverse Modeling for Petrophysically Related Unknown Fields }
We present the inverse modeling results for the configuration in which the hydraulic conductivity and resistivity are created through a known petrophysical relationship. The two resistivity fields are first generated from a log-normal distribution using the Gaussian and exponential kernel, which we call Case 1-1 and Case 1-2, respectively. The covariance scale parameters for the Gaussian (Case 1-1) are 355 m in x and 118 m in z, and for the exponential kernel (Case 1-2) are 447 m in x  and 200 m in z, respectively. After converting the resistivity field to the hydraulic conductivity through the petrophysical relationship (Equation~\ref{eq:phitok}), the groundwater flow is simulated using USGS MODFLOW6 \cite{modflow} to create head and hydraulic conductivity observations at the wells with $1\%$ and $3\%$ noise, respectively. The self-potential model is executed to create electrical potential with $1\%$ noise. For MT observations, MARE2DEM \cite{key2016mare2dem} is used to simulate the MT amplitude and phase.  The modeling and inversion parameters are listed in Table~\ref{table:1}. After conducting the MODFLOW6 simulation, a total of 20 hydraulic head observations and 20 hydraulic conductivity well cores are obtained from four observation wells (as illustrated in Figure~\ref{fig:figure_scheme}). The self-potential model provides 116 electrical potential data from the well and surface, and a total of 5248 MT amplitude and phase data are produced by the MARE2DEM. We run inversions on a computational node with 48 cores and 190 GB RAM. 

\begin{table}[h!]
\caption{Parameter summary used in the application}
\centering
\begin{adjustbox}{width=1\textwidth,center=\textwidth}
\begin{tabular}{c c c}
\hline
 & Description & Application\\
\hline
Geometric Parameters\\
$L_{x}, L_{z}$ & domain length and depth (m) & 2000, 2000 \\
$\Delta x$, $\Delta z$ & grid size (m) & 20, 20\\
\hline
Measurement Error\\
$\sigma_{h}$ & std(error) of head (m) & 0.3\\
$\sigma_{\ln{K}}$ & std(error) of $\ln K$ ($\ln m/d$) & 0.05 \\
$\sigma_{sp}$ & std(error) of self-potential (V) & $10^{-3}$\\
$\sigma_{MT1}$ & std(error) of MT $\log_{10}$amplitude (-) & 0.1\\
$\sigma_{MT2}$ & std(error) of MT phase & 2.0\\
\hline
Geostatistical Parameters\\
$q(x,x')$ & exponential covariance kernel & $\sigma_s^2 \exp(-\frac{r}{\theta_{exp}})$\\
$\theta_{exp}$ & exponential scale parameters in x and z  (m) & [447, 200]\\
$q(x,x')$ & Gaussian covariance kernel & $\sigma_s^2 \exp\left[-\left(\frac{r}{\theta_{gau}}\right)^{2}\right]$\\
$\theta_{gau}$ & Gaussian scale parameters in x and z (m) & [355, 118]\\
$\sigma_{s}$ & prior standard deviation ($\ln m/d$) & 2.0\\
\hline
PCGA Parameters\\
$n_{s_1}$ & The number of principal components for single inversion of $\ln K$ & 200 \\
$n_{s_2}$ & The number of principal components for single inversion of $\ln \rho$ & 200 \\
$n_{s}$ & The number of principal components for joint inversion of $\ln K$ and $\ln \rho$ & 400 \\
\hline
\end{tabular}
\label{table:1}
\end{adjustbox}
\end{table}

\begin{figure}[htp!]
    \centering
    \includegraphics[width=\textwidth]{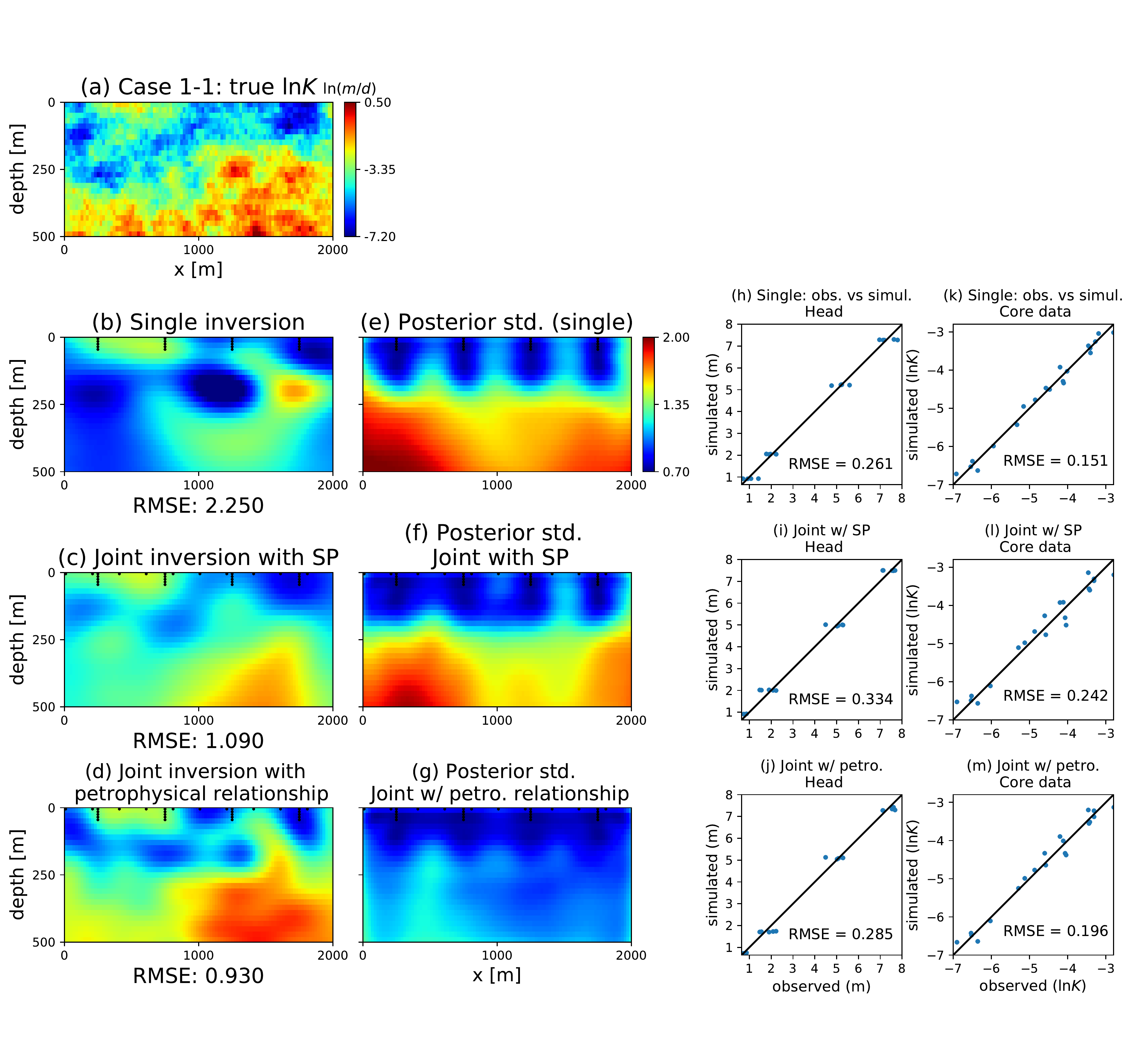}
    \caption{The $\ln{K}$ estimation from single and joint inversion with the petrophysical relationship }
    \label{fig:case1-1_k}
\end{figure}

With this configuration, the ``single`` inversion using  MODFLOW6 and hydrogeological data and the joint data inversion using hydrogeological and geophysics models and datasets are performed to identify and quantify the estimation uncertainty. Figure~\ref{fig:case1-1_k} presents the estimated log-hydraulic conductivity fields from the single and joint inversions up to the depth of 500 m below the ground surface. Beyond the depth of 500 m, the uncertainty reduction from the prior is no longer observed under our configuration thus we decide not to present in this study. The accuracy of the single and joint inversion results is evaluation through the element-wise root-mean-squared-error (RMSE):
\begin{equation}
RMSE = \sqrt{\frac{1}{m} \sum_{i=1}^{m} \left(\ln K_{i}^{est}-\ln K_{i}^{true} \right)^2}
\end{equation}
where $\ln{K}_{i}^{est}$ is the estimated log-hydraulic conductivity in the grid \emph{i}, and $\ln{K}_{i}^{true}$ is the \emph{i}-th true log hydraulic conductivity. The RMSE of log-hydraulic conductivity from single inversion up to the depth of 500 m is 2.25 $\ln(m/d)$ while the RMSE of joint inversion is 1.09 $\ln(m/d)$, which shows a great improvement of 51.6\% in the accuracy compared to the single inversion result.

It is shown that the low hydraulic conductivity regions up to the depth of 100 m are estimated well in both results from the single and joint inversions. Still, the joint inversion produces better results overall in the terms of RMSE since the information from the deeper formation can be acquired from the MT survey. Indeed, the proposed joint inversion approach is able to identify high hydraulic conductivity anomalies below 250 m as shown in the jointly estimated $\ln K$ of Figure~\ref{fig:case1-1_k} (c). In the joint inversion, the use of the self-potential observations links the groundwater flow phenomena to the geophysics so that one can perform reasonably accurate deeper subsurface characterization without any petrophysical relationship.  

Figure~\ref{fig:case1-1_k} (e)-(g) display the posterior standard deviation, i.e., linearized uncertainty, of the single and joint inversion estimates. As reported in many previous studies~\cite{lee2014large}, the inversion using hydrogeological data alone results in high uncertainty away from the observation wells. On the other hand, the proposed joint inversion combining hydrogeological and geophysical data sets leads to lower uncertainty near the observation wells and also a reduction in the uncertainty in the area of the high hydraulic conductivity anomalies below 250 m where the uncertainty of the joint inversion is found to be relatively lower than that from the single inversion. 

The right two columns of Figure~\ref{fig:case1-1_k} shows the scatter plots of observations against simulated hydraulic heads and hydraulic conductivities from the single and joint inversions. The data fittings are reasonably performed within the range of the observation errors specified in Table~\ref{table:1}. The single inversion for $\ln{K}$ requires 4 iterations and takes about 8 seconds for each iteration, and the single inversion for $\ln{\rho}$ needs 6 iterations for simulation and takes 13 minutes for each iteration. The joint inversion of $\ln{K}$ and $\ln{\rho}$ needs 7 iterations to be converged and take 25 minutes for each iteration.

\begin{figure}[htp!]
    \centering
    \includegraphics[width=\textwidth]{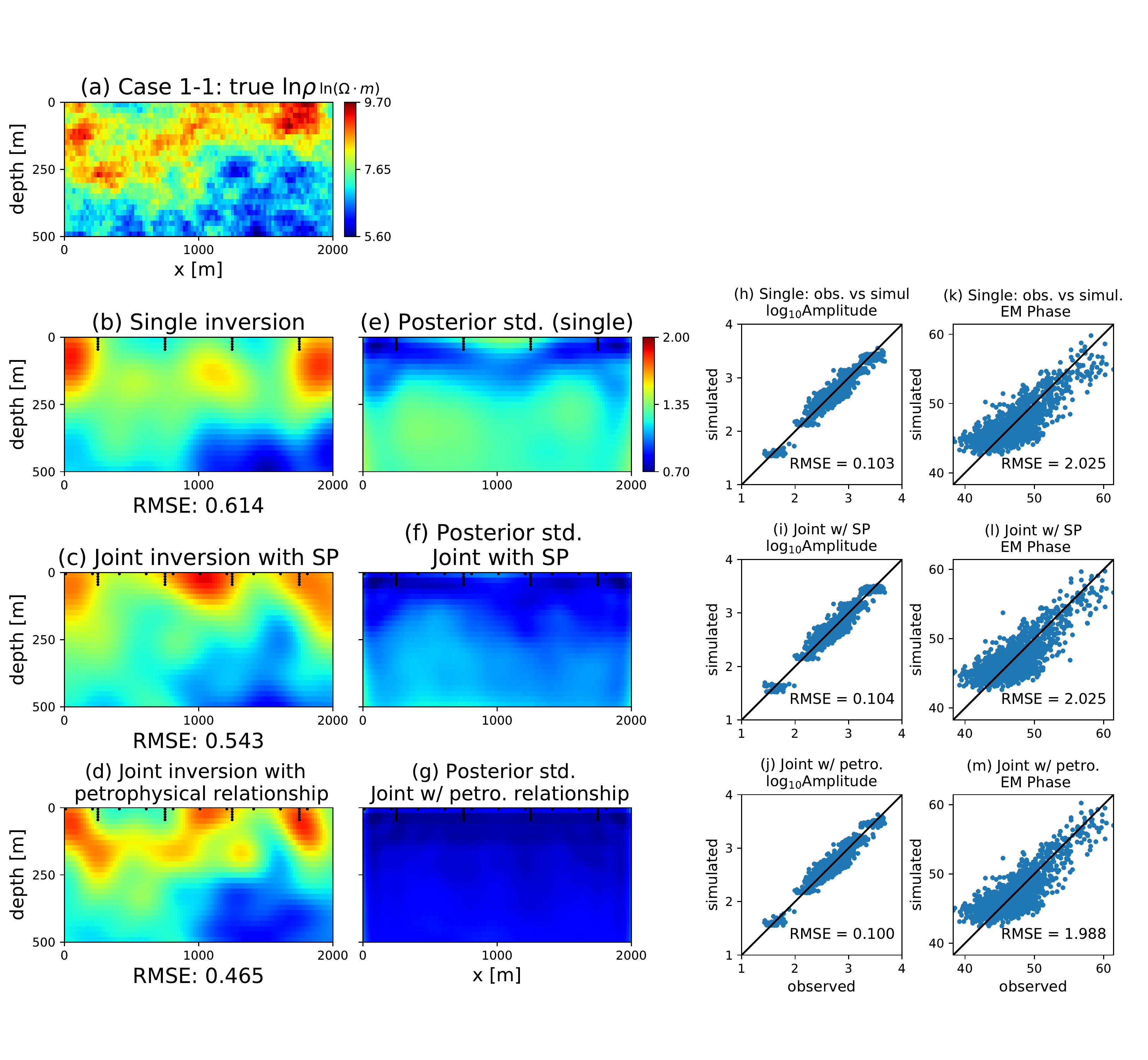}
    \caption{The $\ln{\rho}$ estimation from single and joint inversion with the petrophysical relationship}
    \label{fig:case1-1_res}
\end{figure}

We also perform the resistivity inversion using MT response alone and compare it with the joint data inversion. Figure~\ref{fig:case1-1_res} (b) and (c) show the log-resistivity $\ln{\rho}$ estimate from single and joint inversions. The estimated resistivity fields from both methods are similar to the true value in general while the joint inversion, which additionally utilizes the surface and over 100m in the depth of the observation well data, estimates the resistance field in more detail in shallow areas. In specific, the joint inversion estimates the high resistivity region around \emph{x} = 1000 m below the surface. The RMSEs of estimated $\ln{\rho}$ from single and joint inversions are 0.614 and 0.543, respectively showing that the joint inversion results in a better estimation with 11.6\% improvement in terms of the element-wise RMSE.  
 
Figures \ref{fig:case1-1_res} (e)-(g) present the posterior standard deviation for both cases. The joint inversion estimate has lower uncertainty for the overall region compared to the single inversion. The additional information from the observation wells and the SP can reduce the uncertainty significantly for both shallow (i.e., around \emph{x} = 1000 m) and deeper (i.e., below \emph{z} = 250 m) regions in general. Note that it is observed that the estimation uncertainty is relatively high where the joint inversion fails to identify high resistivity region near \emph{x} = 800 m and \emph{z} = 175 m.

Figures~\ref{fig:case1-1_res} (h)-(m) represents the scatter plots of observations versus simulated MT observation in amplitude and phase. The first two figures (Figure \ref {fig:case1-1_res} (h) and (i)) provide the scatter plots for EM amplitude in log scale from single and joint inversion which show RMSE of 0.103 and 0.104, respectively. The other figures show the comparison between observed and simulated EM phases from single and joint inversions and the RMSE of 2.025 for both inversions.

\subsubsection{Comparison with Joint Inversion using Known Petrophysical Relationship}
An additional experiment is performed in this subsection to investigate how well our proposed method identifies the unknown subsurface aquifer. Since the hydraulic conductivity $\mathbf{K}$ and resistivity $\boldsymbol \rho$ fields are generated with a predefined petrophysical relationship, we utilize this known petrophysical relationship as additional information for estimating the unknown $\mathbf{K}$ and resistivity $\boldsymbol \rho$ along with the hydraulic head and MT data sets. This setting reveals the maximum information the joint inversion using the well-based observations and MT survey can attain and allows us to evaluate the value of the self-potential measurements. 

The petrophysical relationship links the groundwater model with the MT model to update the $\mathbf{K}$ and $\boldsymbol \rho$ fields simultaneously. In the implementation, the additional term is added to the objective function in Eq.~\ref{eq:map}:
\begin{equation}
\frac{1}{2}\left( \mathbf{K}-f_{petro}(1/\ln{ \boldsymbol\rho})  \right)^{\text{T}} \mathbf{R}^{-1}_{petro} \left(\mathbf{K}-f_{petro}(1/\ln{\boldsymbol\rho}) \right) 
\label{eq:obj_PR}
\end{equation}
where $f_{petro}$ represents the hydraulic conductivity estimated with resistivity using the petrophysical relationship described in the previous section~\ref{Subsurface Property and Data Generation} and the diagonal matrix $\mathbf{R}_{petro}$ represents the matrix of the petrophysical relationship errors/residuals. $\mathbf{R}_{petro}$ were assigned based on the error level of estimated hydraulic conductivity.    

Figure~\ref{fig:case1-1_k} (d) shows the estimation of the $K$ field obtained with the petrophysical relationship with the lowest RMSE of 0.930 among the three inversion results from Case 1. Since we assume a perfectly identified petrophysical relationship during the inversion, the result estimates the high and low hydraulic conductivity accurately both in deep and shallow regions. In this specific example, the proposed joint inversion with self-potential data as in Figure~\ref{fig:case1-1_k} (c) yields comparable results to that with perfectly known petrophysical relationship qualitatively and also in terms of RMSE (i.e., RMSE 0.930 vs 1.090). This additional analysis shows that the proposed joint inversion approach via the SP data can perform reasonably accurate subsurface characterization without any petrophysical relationship which may not be accessible or completely wrong in practice. Figure~\ref{fig:case1-1_res} (d) presents the estimation of the $\rho$ field that is also comparable to Figure~\ref{fig:case1-1_res} (c).   
\subsection{Case 2: Independently Generated Unknown Subsurface Fields}
\begin{figure}[htp!]
    \centering
    \includegraphics[width=\textwidth]{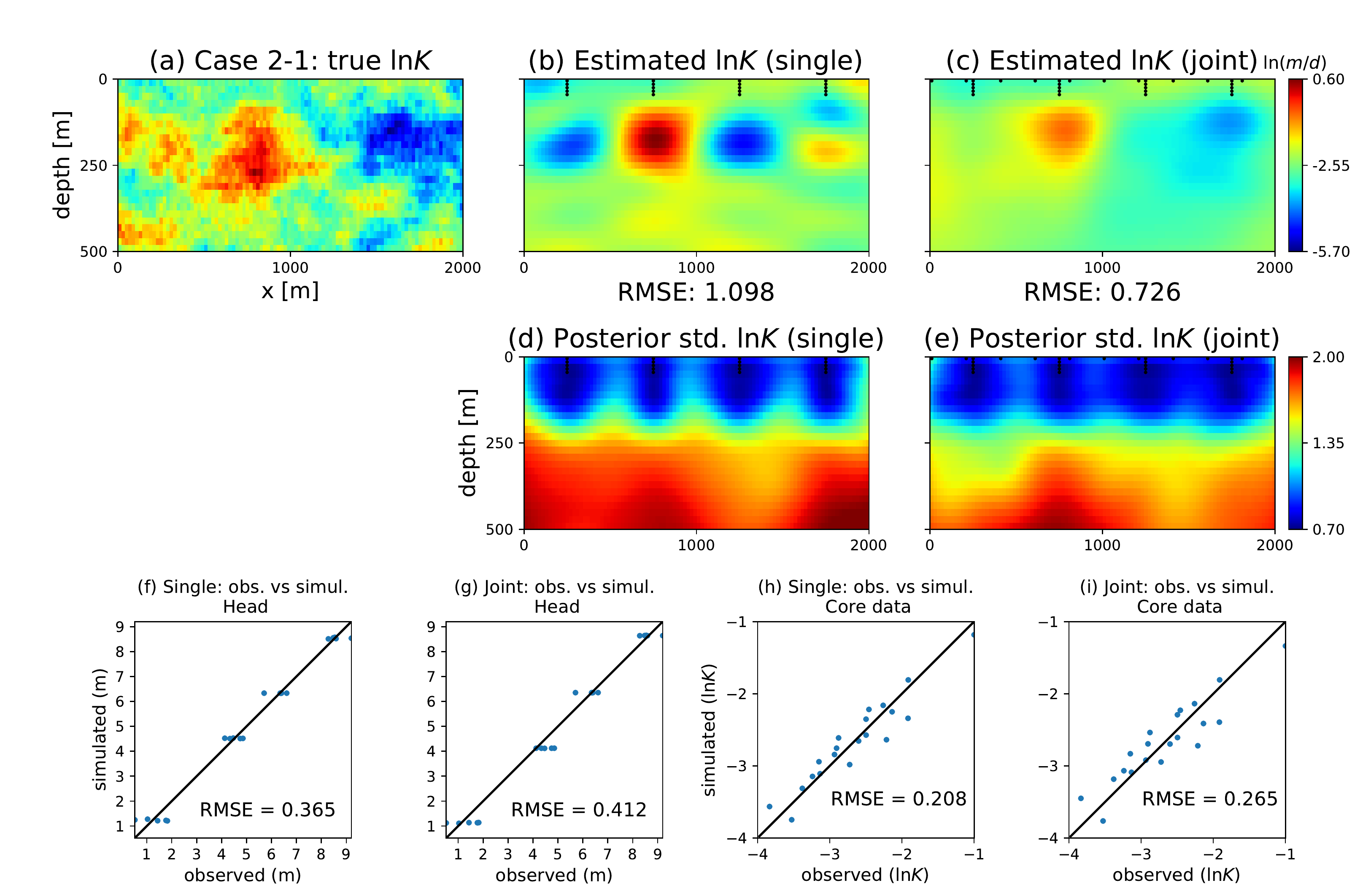}
    \caption{The $\ln{K}$ estimation from single and joint inversion; in Case 2, $K$ and $\rho$ fields are independently generated to ensure no specific relationship between two fields}
    \label{fig:case2-1_k}
\end{figure}

\begin{figure}[htp!]
    \centering
    \includegraphics[width=\textwidth]{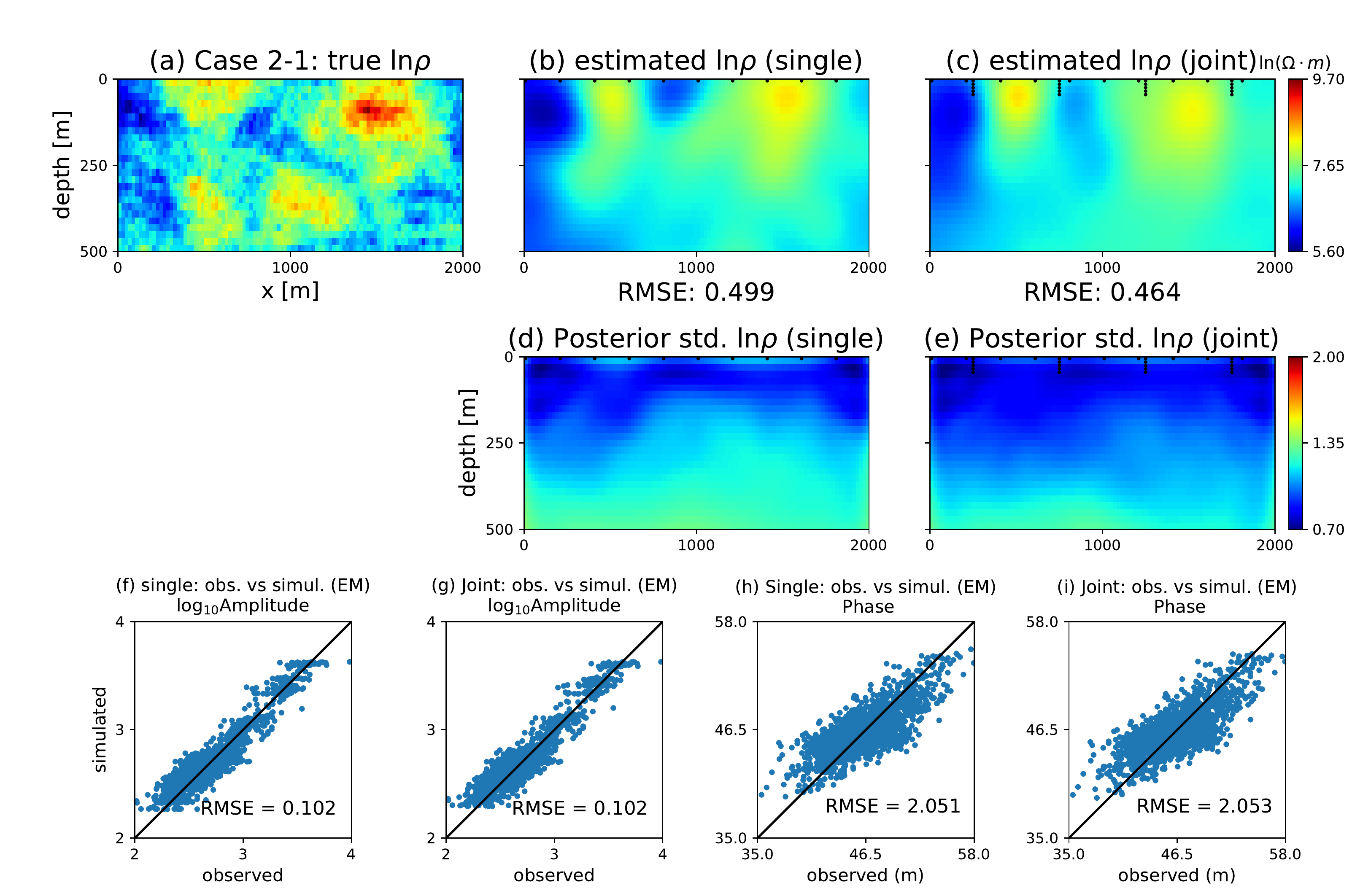}
    \caption{The $\ln{\rho}$ estimation from single and joint inversion in Case 2}
    \label{fig:case2-1_res}
\end{figure}

In this subsection, we examine the applicability of our proposed method to the synthetic true hydraulic conductivity and resistivity fields that are created without any petrophysical relationship. The true hydraulic conductivity and resistivity fields are generated independently with the previously used geostatistical models and their associated parameters listed in Table~\ref{table:1}. The estimated log hydraulic conductivity and resistivity fields are shown in Figure~\ref{fig:case2-1_k} (a) and Figure~\ref{fig:case2-1_res} (a), respectively. 

Figures~\ref{fig:case2-1_k} (b)-(c) represent the estimated hydraulic conductivity field from the single and joint inverse modelings. The RMSE of single and joint inversions are 1.353 and 0.726 $\ln(m/d)$, respectively, which indicates a 46.4\% improvement in RMSE. The joint inversion as in Figures~\ref{fig:case2-1_k} accurately identifies high and low conductivity regions up to the depth of 250 m while the single inversion could not produce reliable results beyond the depth of 100 m where the observation well penetration ends and wrongly estimated high-low conductivity areas are artifacts from the inversion showing that one would not resort in the values outside the observation network, i.e., ``extrapolation''. The proposed approach identifies the hydraulic conductivity field well regardless of the existence of any underlying petrophysical relationship.

The posterior uncertainty plots in Figure~\ref{fig:case2-1_k} (d)-(e) illustrate overall uncertainty reduction nearby observation wells while the joint inversion result produces lower uncertainty between observation wells due to the additional measurements from SP and MT surveys. The joint inversion result identifies the high-low conductivity variation below the observation wells due to the coverage of the MT survey and the uncertainty map confirms the increased accuracy of the estimation up to a depth of 400 m. The scatter plots of data fitting are presented in Figure~\ref{fig:case2-1_k} showing the data are fitted reasonably within the predefined measurement error levels. The cR/Q2 criteria~\cite{kitanidis1991orthonormal}, which is an optimal structure (hyper) parameter selection method in the Bayesian framework, was also performed to check the validity of the parameter selection in this study. 

Figure \ref{fig:case2-1_res} represents the estimation of $\ln{\rho}$ by the depth of 500 m. The first figure shows the true $\ln{\rho}$ field, and the other figures show the estimated $\ln{\rho}$ by single and joint inversions, respectively. The RMSE of estimated $\ln{\rho}$ from single inversion is 0.499 and 0.464, respectively. The $\ln{\rho}$ joint inversion of case 2-1 improved the element-wise RMSE by 7.1$\%$  without the one-to-one petrophysical relationship between unknowns. 


The posterior standard deviation plots in Figure \ref{fig:case2-1_res} explain that the joint inversion has lower uncertainty for the overall domain and can have a high-resolution interval between the depth of 0 m and 250 m. With both methods, the PCGA in case 2 can also estimate $\ln{\rho}$ fields that are similar to the true value within the range between 0 and 500 m. As in case 1, the joint inversion of case 2 can more accurately reproduce the resistivity field of which the depth near the surface and observation wells.

\subsection{Additional Tests}
\begin{figure}[htp!]
    \centering
    \includegraphics[scale = 0.5]{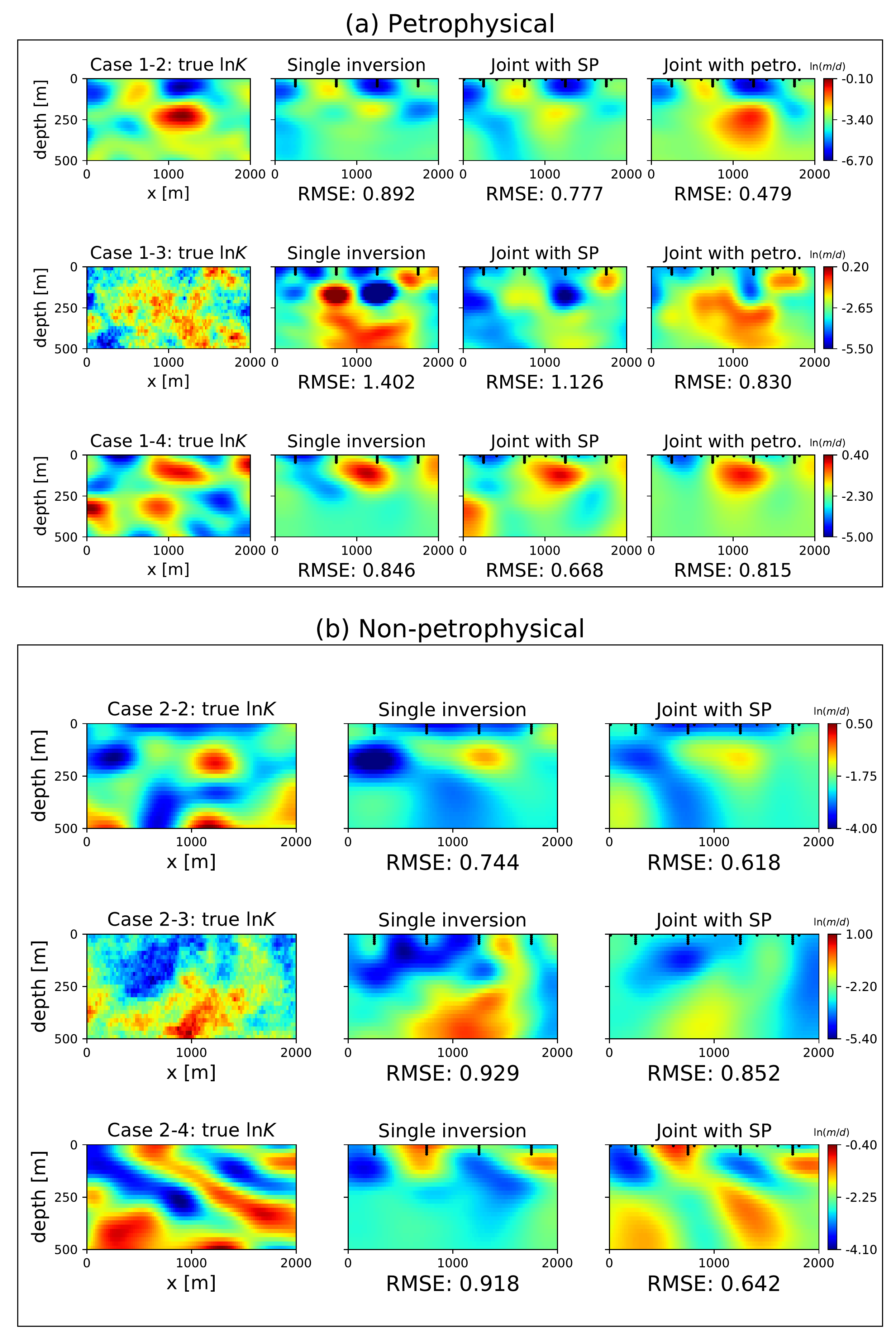}
    \caption{The $\ln{K}$ estimations for (a) Case 1: unknown $K$ and $\rho$ fields generated from the petrophysical relationship and (b) Case 2: independently generated $K$ and $\rho$ fields}
    \label{fig:cases_k}
\end{figure}

\begin{figure}[htp!]
    \centering
    \includegraphics[width=\textwidth]{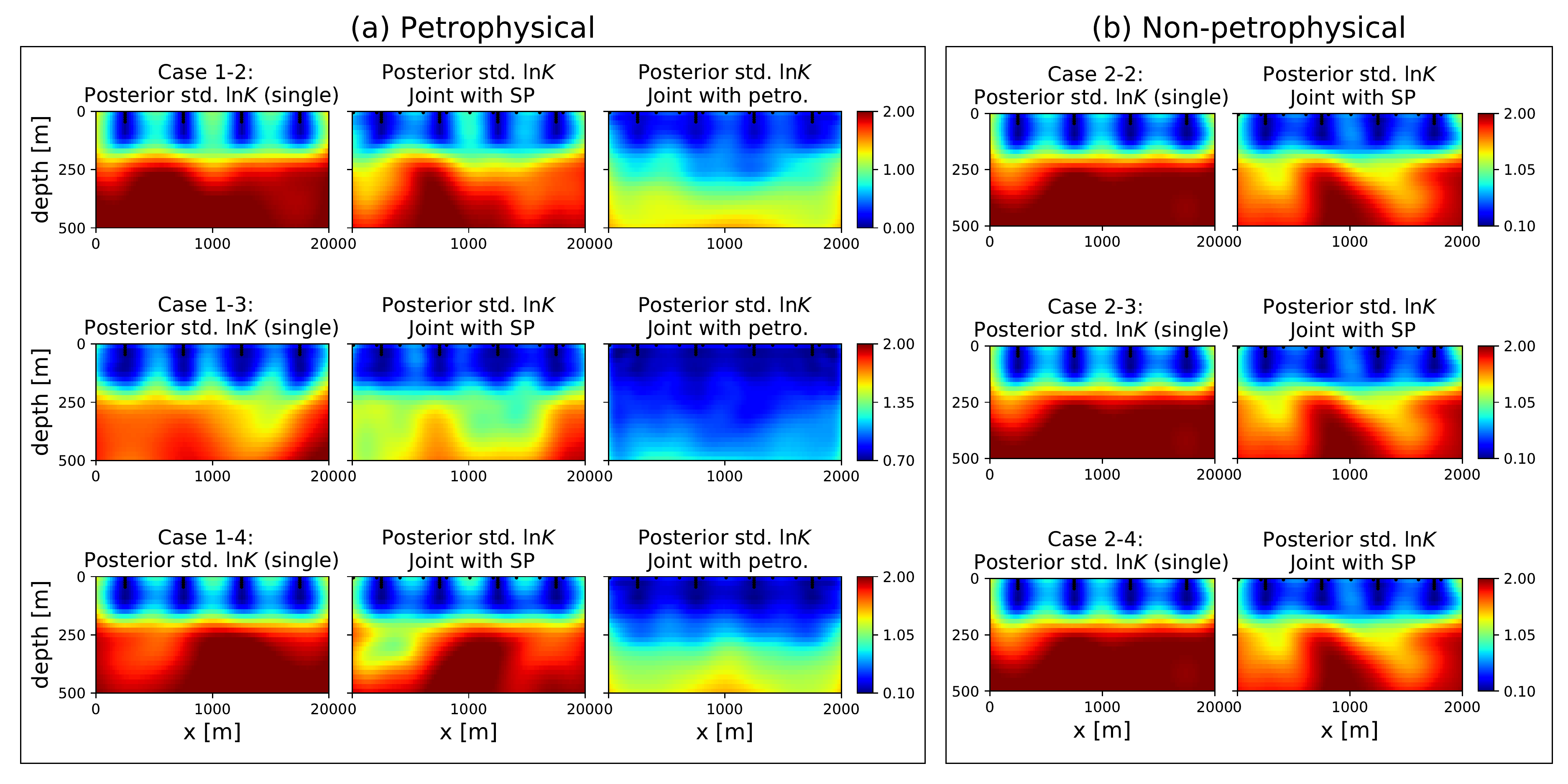}
    \caption{The Uncertainty of $\ln{K}$ estimations for (a) Case 1: unknown $K$ and $\rho$ fields generated from the petrophysical relationship and (b) Case 2: independently generated $K$ and $\rho$ fields}
    \label{fig:cases_uns}
\end{figure}

We generate additional six hydraulic conductivity and resistivity fields to validate our proposed method further. The first three hydraulic conductivity and resistivity fields are created with a one-to-one petrophysical relationship (Case 1) and the other three fields are generated independently (Case 2) as in the previous sections. For each case, three examples in total are created, one using the exponential kernel as in previous results of Case 1-1 and Case 2-1 and the other two examples generated by the Gaussian kernel in order to test different covariance models. Figures~\ref{fig:cases_k} and~\ref{fig:cases_uns} show the $\ln{K}$ estimates and their corresponding uncertainty results, respectively. As observed in the previous sections, the hydraulic conductivity estimate is consistently better than those from the single inversion and even comparable to the joint inversion results when we can access the ``true'' petrophysical relationship.

For comparison across all the tests with different ranges of hydraulic conductivity and resistivity, we also compute the normalized RMSE (NRMSE) as:
\begin{equation}
\textrm{NRMSE} = \frac{\textrm{RMSE}}{y_{max}-y_{min}}
\end{equation}
where $y_{max}$ is the maximum value of the observations and $y_{min}$ is the minimum value of the observations. The RMSE and NRMSE of $\ln{K}$ estimation and the (relative) improvement of the joint inversion results with respect to those from the single inversion are shown in Table~\ref{table:2}. The average improvements from our proposed joint inversion approach for Case 1 and Case 2 are 25.8\% and 25.4 \%, respectively.

\begin{table}[h!]
\caption{RMSE and NRMSE of $\ln{K}$ estimation}
\begin{tabularx}{0.95\textwidth} { 
   >{\centering\arraybackslash}X
   >{\centering\arraybackslash}X 
   >{\centering\arraybackslash}X 
   >{\centering\arraybackslash}X 
   >{\centering\arraybackslash}X }
\toprule
 \multicolumn{5}{c}{RMSE $\left(\ln{(m/d)}\right)$ / NRMSE (-)} \\
 \hline
 Cases& Single Inv.& \begin{tabular}{@{}c@{}} Joint Inv.\\ with SP \end{tabular}  & \begin{tabular}{@{}c@{}}Joint Inv.\\ with petrophysical \end{tabular} & \begin{tabular}{@{}c@{}} Improvement \\ (\%) \end{tabular}   \\
 \hline
 Case 1-1   & 2.25 / 0.29 & 1.09 / 0.14 & 0.93 / 0.12 & 51.6\\
 Case 1-2   & 0.89 / 0.14 & 0.78 / 0.12 & 0.48 / 0.07 & 13.0\\
 Case 1-3   & 1.40 / 0.25 & 1.13 / 0.20 & 0.83 / 0.15 & 19.7\\
 Case 1-4   & 0.85 / 0.16 & 0.67 / 0.12 & 0.82 / 0.15 & 20.9\\
 \hline
 Case 2-1 & 1.10 / 0.18 & 0.73 / 0.12 & - & 33.9\\
 Case 2-2 & 0.74 / 0.17 & 0.62 / 0.14 & - & 17.0\\
 Case 2-3 & 0.93 / 0.15 & 0.85 / 0.13 & - & 8.2\\
 Case 2-4 & 0.92 / 0.25 & 0.64 / 0.17 & - & 30.0 \\
\bottomrule
\end{tabularx}
\label{table:2}
\end{table} 

\begin{figure}[htp!]
    \centering
    \includegraphics[scale = 0.5]{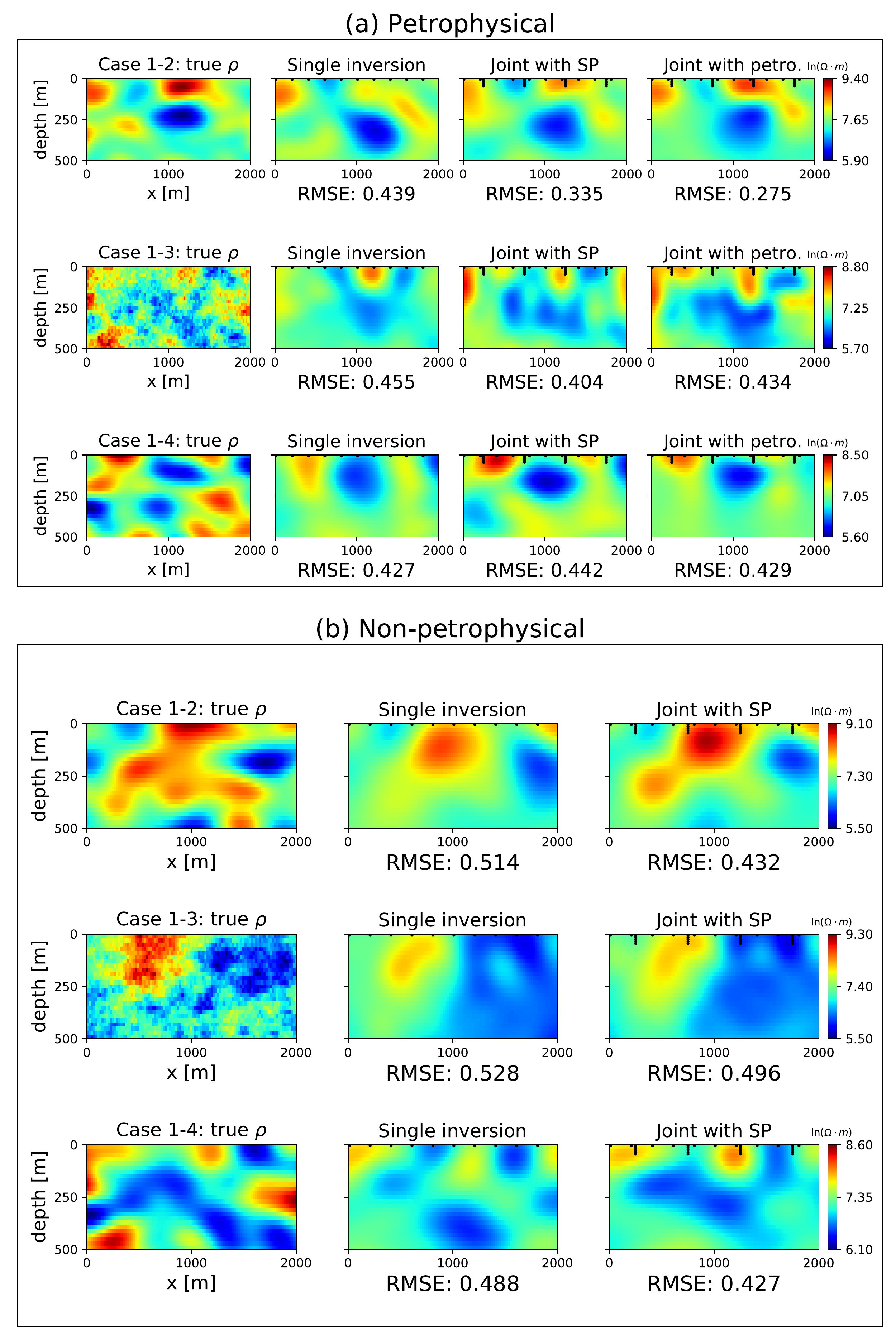}
    \caption{The $\ln{\rho}$ estimations for (a) Case 1: unknown $K$ and $\rho$ fields generated from the petrophysical relationship and (b) Case 2: independently generated $K$ and $\rho$ fields}
    \label{fig:cases_res}
\end{figure}

\begin{figure}[htp!]
    \centering
    \includegraphics[width=\textwidth]{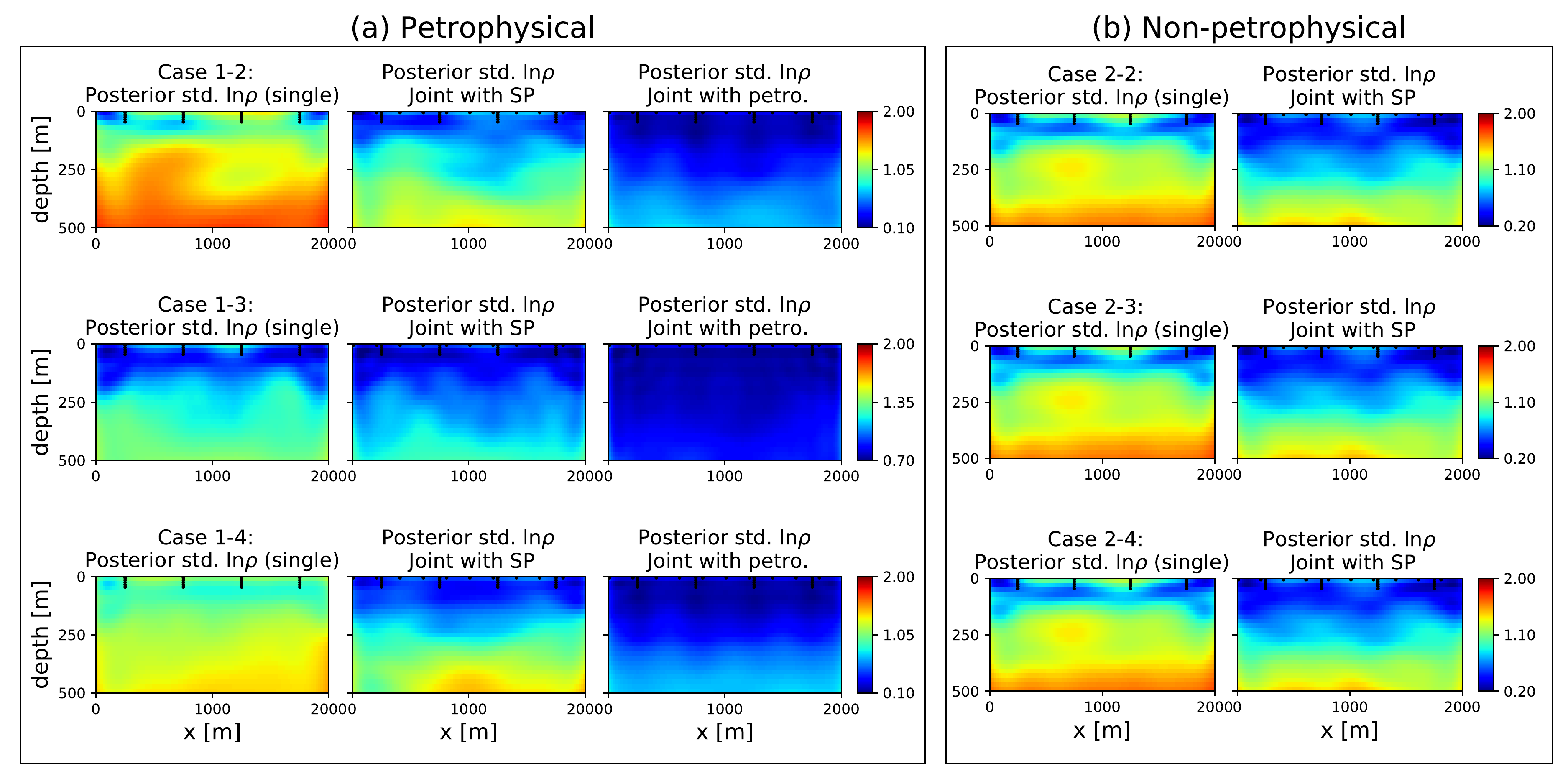}
    \caption{The uncertainty of $\ln{\rho}$ estimations for (a) Case 1: unknown $K$ and $\rho$ fields generated from the petrophysical relationship and (b) Case 2: independently generated $K$ and $\rho$ fields}
    \label{fig:cases_uns_res}
\end{figure}

Figure~\ref{fig:cases_res} presents the estimates of $\ln{\rho}$ from the single MT and joint inversions. Overall, the proposed joint inversion strategy produces slightly better results than those from the single MT inversion due to the additional information from hydrogeological data sets and SP measurements while both single and joint inversions provide reasonable estimations. This is because the MT survey can image deeper subsurface and the drastic improvement observed in the hydraulic conductivity estimation is not expected in the resistivity estimation, especially in the deeper region. 
Figure~\ref{fig:cases_uns_res} represents the uncertainty plots of the resistivity estimation for petrophysically related and independently created unknown fields. Because of the broader range of the MT survey, the uncertainty reduction in the resistivity estimates is more widely observed than the uncertainty reduction in the hydraulic conductivity estimation as in Figure~\ref{fig:cases_uns}. Still, the proposed joint inversion provides uncertainty reduction deeper through additional information from observation wells and SP data. Note that the use of the true petrophysical relationship (Figure~\ref{fig:cases_uns} (a)) can reduce the estimation uncertainty further because we constrain the relationship at every location of the aquifer during the inversion. Of course, the ``true'' petrophysical relationship will not be available in practice and all the numerical tests illustrate the effectiveness of SP-based joint inversion without any petrophysical assumptions. 

\clearpage

\section{Concluding Remarks}
\label{sec:conclusion}
This paper examines a new joint inversion strategy utilizing self-potential data, hydraulic head, and MT data set for estimating hydraulic conductivity and resistivity of a deep aquifer formation that exceeds typical hydrogeological investigations. The joint inversion only needs the forward models of MT, self-potential, and groundwater flow with associated data sets without assuming any petrophysical relationship between hydrogeological variables. The need for such a relationship is eliminated for the joint hydrogeophysical inversion by using the self-potential equation utilizing self-potential surveys to link hydraulic conductivity and resistivity. The joint inversion estimates hydraulic conductivity and electrical resistivity fields simultaneously, by using hydraulic head and core data from observation wells, self-potential data from surface acquisition and the wells, and MT data from MT receivers.

Several subsurface examples for different spatial models are tested to illustrate the effectiveness of the proposed method.  For each experimental example, the joint inversion results are compared against the single inversion results in which hydraulic conductivity or resistivity are estimated independently by using a single forward model for the former, and associated observations for the latter. 

The subsurface examples are considered for two cases, one with petrophysically related and the other with independently generated hydraulic conductivity and resistivity fields. It is shown that regardless of the use of the petrophysical relationship in the data generation, the joint inversion consistently produces estimations close to the true fields up to the depth of 500 m, whereas the single inversion is limited to identifying the subsurface properties only around the wells for hydraulic conductivity and up to the depth of 250 m for the resistivity. For the hydraulic conductivity estimation, the posterior standard deviation in the single inversions shows low uncertainty zones only nearby observation wells and abrupt increases in uncertainty below 250 m. On the other hand, the joint inversions produce wide and deeper low uncertainty zones across and below the wells. For the resistivity estimation, the joint inversion shows a deeper and larger uncertainty reduction than the single inversion. Additionally, for the case where the petrophysical relationship is used for the data generation, the joint inversion shows close agreement with results that are obtained by the proposed joint approach utilizing the self-potential survey. 

To achieve the results presented in this paper, we use a computing workstation equipped with 48 cores and 190 GM RAM within a high-performance computing cluster for conducting the inverse modeling. The numbers of principal components for the PCGA inversion were set to 200 and 400 for single and joint inversion (200 each for hydraulic conductivity and resistivity), respectively. The single $\ln{K}$ estimation inversions using the groundwater equation required 4 iterations on average and took about 32 seconds to converge, and the single $\ln{\rho}$ inversions using the MT forward solver needed 5 iterations and took 65 minutes in average. The single inversion executed about 200 forward model runs for each iteration to be converged. The joint inversion of $\ln{K}$ and $\ln{\rho}$ took about 150 minutes with 6 iterations on average with the number of joint forward model runs, which consists of groundwater, SP, and MT models, to be about 200 times for each iteration. This illustrates the computational efficiency of the current proposed approach for deep aquifer characterization using joint data sets, which can scale suitably to three-dimensional applications~\cite{lee2016scalable}.

The current paper presents and illustrates a potential strategy to address the current limitation of unknown or uncertain petrophysical relationships for deep aquifer characterization, using a combination of hydrogeophysical and hydrogeological data sets. Additional validation studies using field data are needed to validate the proposed method further.
It is important to note that, even though this paper demonstrates a combination of MT, SP, and hydrogeological data to prove the concept of the proposed methodology, the methodological framework is transferable to other combinations of data types. To this end, different combinations of geophysical data sets could be explored, such as 3D electrical resistivity tomography data (ERT) instead of MT data, or active seismic data and/or ambient seismic noise data combined with slightly different parameterizations.

One can reproduce the results reported in the paper using the code \url{https://github.com/yhseo0321/Inverse_Modeling}. The link to the PCGA inversion code is available in~\url{https://github.com/jonghyunharrylee/pyPCGA}. 

\section*{Acknowledgments}
The work is supported by the National Science Foundation, under Office of Advanced Cyber Infrastructure Award Number 2005259. The technical support and advanced computing resources from the University of Hawaii Information Technology Services – Cyberinfrastructure, funded in part by the National Science Foundation CC* awards \# 2201428 and \# 2232862 are gratefully acknowledged. The code used in the paper can be found in~\url{https://github.com/yhseo0321/Inverse_Modeling} and~\url{https://github.com/jonghyunharrylee/pyPCGA}

\bibliographystyle{unsrt}  
\bibliography{references}

\end{document}